\documentclass{Interspeech2024}

\interspeechcameraready
\title{On Calibration of Speech Classification Models: Insights from Energy-Based Model Investigations}

\name{Yaqian}{Hao$^\dagger$}
\name{Chenguang}{Hu$^\dagger$}
\name{Yingying}{Gao}
\name{Shilei}{Zhang$^*$}
\name{Junlan}{Feng$^*$}

\address{
  China Mobile Research Institute, Beijing, China}
 \email{\{haoyaqian, huchenguang,  gaoyingying, zhangshilei, fengjunlan\}@chinamobile.com}


\begin{document}

\maketitle

\begin{abstract}
    For speech classification tasks, deep learning models often achieve high accuracy but exhibit shortcomings in calibration, manifesting as classifiers exhibiting overconfidence. The significance of calibration lies in its critical role in guaranteeing the reliability of decision-making within deep learning systems. This study explores the effectiveness of Energy-Based Models (EBMs) in calibrating confidence for speech classification tasks by training a joint EBM integrating a discriminative and a generative model, thereby enhancing the classifier’s calibration and mitigating overconfidence. Experimental evaluations conducted on three speech classification tasks specifically: age, emotion, and language recognition. Our findings highlight the competitive performance of EBMs in calibrating the speech classification models. This research emphasizes the potential of EBMs in speech classification tasks, demonstrating their ability to enhance calibration without sacrificing accuracy.
\end{abstract}

\renewcommand{\thefootnote}{\fnsymbol{footnote}}
\footnote{Corresponding Author.}
\footnote{Equal Contribution.}
\renewcommand{\thefootnote}{\arabic{footnote}}

\section{Introduction}

\label{into}
Despite the impressive performance of deep learning models in speech classification~\cite{rajendran2021language}, issues such as overconfidence, calibration errors, and uncertainty estimation may hinder their reliability and generalization in real-world scenarios~\cite{yu2011calibration,guo2017calibration,gitman2023confidence}. Confidence calibration in these models poses a significant challenge~\cite{yu2011calibration,huang2021context,chouimportance}. For example, in speech emotion recognition (SER) systems, the inherent uncertainty in modeling emotions affects the trustworthiness of the model's predictions~\cite{chouimportance}.
Overconfidence and underconfidence can indicate suboptimal calibration, leading to false positives or missed opportunities~\cite{guo2017calibration,gitman2023confidence,kubik2022underconfidence}. The current state of research in speech classification models often overlooks the issue of confidence calibration, resulting in a lack of reliable methods and leading to uncertainty in predictions. The current state of confidence calibration in speech classification underscores a lack of reliable methods, fostering uncertainty and mistrust in model predictions.

Undoubtedly, ensuring a well-calibrated confidence measure in a classification model is crucial for accurate predictions. Therefore, developing methodologies to adjust the predictions of a speech classification model is essential, balancing calibration with performance. 
Existing techniques, like Temperature Scaling and Vector Scaling, typically rescale the posterior distributions of classifier predictions~\cite{guo2017calibration,wang2021confident,wang2023calibrating}. However, these methods need post-processing adjustments and require a persistent development set with enough samples. Alternatively, adjusting calibration during the model training process, such as using confidence regularization, offers another approach ~\cite{pereyra2017regularizing,jung2020posterior}.

Recently, the effectiveness of EBMs in achieving enhanced model calibration has been demonstrated~\cite{grathwohl2019your}, wherein the joint training process incorporates both discriminative and generative models. 
The EBMs characterize the relationship between density of input data and model energy, enabling predictions based on energy minimization. While this flexibility is advantageous, the training process of energy models involves intricate adjustments, making it a challenging endeavor~\cite{wang2021confident,song2021train}. Following ~\cite{grathwohl2019your}'s work, ~\cite{yang2021jem++,yang2023towards} has intricately improved the training process of EBMs, substantially boosting both training efficiency and ultimate performance. 
Despite their effectiveness in computer vision, EBMs' potential for calibrating speech classification models remains untapped.
In this paper, we explore the effectiveness of EBMs in enhancing the calibration of speech classification models. Through experiments on three distinct speech classification tasks, we compare EBMs with traditional softmax-based models. Results reveal that EBMs achieve an average reduction of 7.787\% in Expected Calibration Error (ECE) across the tasks, indicating improved calibration. Additionally, Negative Log-Likelihood (NLL) shows an average reduction of 0.172, indicating enhanced model fitting to observed data and more accurate probability predictions. 
Furthermore, we compared EBMs with other calibration methods such as Temperature Scaling and Logistic Scaling, and the results demonstrate that EBMs exhibit a significantly greater reduction in overconfidence compared to  these post-processing methods.

The key contributions of this paper are summarized as follows:  
\begin{enumerate}
\item We introduce joint EBMs in speech classification tasks to improve calibration by modeling the energy function with a deep neural network, maintaining accuracy while enhancing reliability. This joint energy model optimizes not only for classification tasks but also learns the underlying probability distribution within the data, resulting in improved calibration observed through the model's cautious decision-making when encountering inputs deviating from the training data distribution.
    \item 
    We assess the performance of EBMs across three speech tasks and datasets, specifically targeting language, emotion, and age recognition. This evaluation demonstrates that EBMs can significantly reduce ECE without compromising model accuracy, and mitigate overconfidence issues in speech classification models.
 \item  We conduct comparative analyses with other calibration methods, and explore model training dynamics and confidence distributions to address model overconfidence. 
Specifically, our results show that EBMs outperform other post-processing methods in achieving effective calibration without requiring additional auxiliary datasets.
\end{enumerate}
\section{Method}
\label{section:method}
\subsection{Energy-based Models}
The fundamental principle underlying an EBM is to construct a function $E(\mathbf{x}):R^D \rightarrow R$ that maps each point $\mathbf{}$ in the input space to a singular, non-probabilistic scalar referred to as the energy \cite{lecun2006tutorial,ou2024energy}. This scalar, denoted by $E(\mathbf{x})$, is a key component in the Gibbs distribution, allowing the derivation of a probability density $p(\mathbf{x})$. The relationship is formalized as follows:
\begin{equation}
p_\theta (\mathbf{x}) = \frac{\exp\left(-E_\theta(\mathbf{x})/T\right) }{Z(\theta)},
\end{equation}
where $E_\theta(\mathbf{x})$ representing the energy, is a nonlinear regression function parameterized by $\theta$,  $T$ refers to the temperature parameter, and $Z(\theta)$ signifies the normalizing constant, also known as the partition function:
 $$Z(\theta) = \int_\mathbf{x} \exp\left(-E_\theta (\mathbf{x}) /T\right) dx.$$
 \subsection{Energy-based Classifier}
The EBM exhibits an intrinsic association with contemporary machine learning, particularly  discriminative neural classifier \cite{grathwohl2019your,liu2020energy} $f_\theta(\mathbf{x}) :\mathbb{R}^D \rightarrow \mathbb{R}^K$. This classifier assigns logits to each class for a given input $\mathbf{x}$ using the softmax function: \begin{equation}
p_\theta(y | \mathbf{x}) = \frac{\exp(f_\theta(\mathbf{x})[y]/T)}{\sum_i^K \exp(f_\theta(\mathbf{x})[i]/T)},
\end{equation}
where $f_\theta(\mathbf{x})[y]$ denotes the logit associated with the $y$-th class label within the output of $f_\theta(\mathbf{x})$.

This connection allows us to view the discriminative classifier $f(\mathbf{x})$ as an energy function in the EBM framework $E_\theta(\mathbf{x}, y) = -f_\theta(\mathbf{x})[y]/T$. Consequently, the Helmholtz free energy function $E(\mathbf{x}; f)$ for a given data point $\mathbf{x} \in \mathbb{R}^D$ can be represented as the negative logarithm of the partition function:
\begin{equation} 
E_\theta(\mathbf{x}; f) = -T \cdot \log \sum_i^K \exp(f_\theta(\mathbf{x})[i]/T). \nonumber
\end{equation}
Additionally, the logits from $f(\mathbf{x})$ enable the definition of an EBM for the joint distribution of data points $\mathbf{x}$ and labels  $y$:
\begin{equation}
p_\theta(\mathbf{x}, y) = \frac{\exp(-E_\theta(x,y))}{\int_x \sum_i^K \exp(-E_\theta(x,y)}dx)=\frac{\exp(f_\theta(\mathbf{x})[y]/T)}{Z(\theta)}.
\end{equation}
Marginalizing over $y$ provides an unnormalized density model for $\mathbf{x}$:
\begin{equation}
p_\theta(\mathbf{x}) = \sum_y p_\theta(\mathbf{x}, y) = \sum_y \frac{\exp(f_\theta(\mathbf{x})[y]/T)}{Z(\theta)},
\end{equation}
which is precisely the definition of  EBM.
This reinterpretation underscores the intrinsic compatibility between the softmax classifier and the EBM, offering a unified perspective on their shared principles.
 
\subsection{Optimization}
We employ a joint model, integrating an energy-based classifier and a generative model, wherein the EBM is trained to learn the energy function that best captures the data distribution~\cite{grathwohl2019your,song2021train}.
The objective function is as following: 
\begin{equation}\label{joint-prob}
\log p_\theta(\mathbf{x}, y) = \log p_\theta(y|\mathbf{x}) + \log  p_\theta(\mathbf{x}),
\end{equation}
which represents the logarithm of joint distribution of data and labels. 
The conditional distribution $p_\theta(y|\mathbf{x})$ signifies the softmax classification model, while $p_\theta(\mathbf{x})$ captures the marginal data distribution. The loss function is then aligned with the logarithm of the likelihood, as explained in the following:
\begin{footnotesize}
\begin{align}
\label{joint-log}
\log p_\theta(\mathbf{x},y) &=\log p_\theta(y|\mathbf{x}) + \log  p_\theta(\mathbf{x}) \nonumber \\
=& \log \frac{\exp(f_\theta(\mathbf{x})[y]/T)}{\sum_i \exp(f_\theta(\mathbf{x})[i]/T)} +\log \sum_y\frac{\exp(f_\theta(\mathbf{x})[y])/T}{Z_\theta}.
\end{align}
\end{footnotesize}
The derivative of the first term in the Eq.\eqref{joint-log} is relatively straightforward, representing the loss function for training the classifier. The derivative of the second term is:
\begin{footnotesize}
\begin{align}\label{grad}
\nabla_\theta \log p(\mathbf{x})\hspace{-1.5cm}& \nonumber \\
&= \nabla_\theta \log \sum_y \exp(\frac{f_\theta(\mathbf{x})[y]}{T}) - \nabla_\theta \log Z(\theta)\nonumber \\
&= \nabla_\theta \log \sum_y \exp(\frac{ f_\theta(\mathbf{x})[y]}{T})-\mathbb{E}_{x \sim p_\theta(x)}[\log \sum_y \exp(\frac{ f_\theta(\mathbf{x})[y]}{T})] \nonumber \\
&=-\nabla_\theta E_\theta(\mathbf{x}) + \mathbb{E}_{x \sim p_\theta(x)}[\nabla_\theta E_\theta(\mathbf{x}) ].    
\end{align}\end{footnotesize}
By employing a one-sample Monte Carlo estimate $\nabla_\theta \log Z_\theta \sim -\nabla_\theta E_\theta(\tilde{x})$,  where $\tilde{x}$ is sampled from the EBM's distribution $p_\theta(x)$.


\subsection{SGLD-Based Training Method}
According to Eq.~\eqref{grad}, we utilize Langevin MCMC for sampling from $p_\theta(\mathbf{x})$ to train the EBMs~\cite{song2021train}.  Stochastic Gradient Langevin Dynamics (SGLD) is a dynamic optimization algorithm that combines the principles of stochastic gradient descent with Langevin dynamics. To initiate the Langevin sampling process, we begin by drawing an initial sample $x_0$ from a straightforward prior distribution. Subsequently, we simulate an overdamped Langevin diffusion process for $K$ steps, employing a positive step size $\epsilon > 0$. The iteration for each step $k = 0, 1, \ldots, K - 1$ is expressed as:
\begin{eqnarray}
    x_{k+1} &=& x_k + \frac{\epsilon^2}{2}  \nabla_{x_k} \log p_\theta(x_k)   + \epsilon z_k\nonumber 
     \\
     &= &   x_k - \frac{\epsilon^2}{2}  \nabla_{x_k} E_\theta(x_k)  + \epsilon z_k,
\end{eqnarray}
where $\nabla_{x_k} \log p_\theta(x_k)$ represents the gradient of the log probability with respect to $x_k$, and $z_k$ is a random noise term. Notably, as $\epsilon \rightarrow 0$ and $K \rightarrow \infty$, the final sample $x_K$ converges to a distribution that matches $p_\theta(x)$ under certain regularity conditions.


\subsection{Evaluation Metrics}
\textbf{Expected Calibration Error  in Classification.} 
A calibrated classifier aligns confidence with accuracy \cite{mortier2023calibration}.  ECE quantifies calibration by binning predictions and measuring the difference between expected confidence and accuracy.
Mathematically, ECE is expressed as:
\begin{equation}
\small
ECE = \sum_{b=1}^{B} \frac{\lvert B_b \rvert}{N} \lvert \text{acc}(B_b) - \text{conf}(B_b) \rvert,
\end{equation}
where $B$ is the number of bins, $B_b$ represents the $b\)-th bin, $\lvert B_b \rvert$ is the number of samples in bin $B_b$, $N$ is the total number of samples, $\text{acc}(B_b)$ is the average accuracy in bin $B_b$, and $\text{conf}(B_b)$ is the average confidence in bin $B_b$.\\
\textbf{Negative Log-Likelihood in Classification.}
NLL is a key metric for assessing a classification model's calibration. It measures the agreement between predicted probabilities and actual labels by computing the logarithm of the predicted probability assigned to the true label for each sample:
\begin{equation} \text{NLL} = -\frac{1}{N} \sum_{i=1}^{N} \log(P(\hat{y_i})|x_i),\end{equation} 
where \( N \) is the total number of samples, \( \hat{y_i} \) represents the true label of the \( i \)-th sample, and \( P(\hat{y_i}) \) denotes the predicted probability associated with the true label. A lower NLL indicates better calibration, signifying that the model's predicted probabilities closely match the actual outcomes.
In this study, we concentrate on calibration performance, aiming to demonstrate that incorporating EBMs improves confidence calibration without impacting the model’s classification effectiveness. Consequently, we use accuracy as the primary metric to affirm the preservation of core classification capabilities.






\section{Experiments}
\label{section:exp}
\subsection{Datasets}
Multiple datasets will be used in these experiments, with the speech sampled at 16 kHz. The duration of training data for each task and the data split of those datasets is listed in Table~\ref{tab:data}. AP17-OLR~\cite{wang2016ap16}, CASIA~\cite{li2017cheavd} and VoxCeleb-Enrichment~\cite{hechmi2021voxceleb} datasets are used in our experiments for language, emotion and the age group classification respectively.
\vspace{-0.1cm}
\begin{table}[htbp]
\centering
\caption{Dataset}\vspace{-0.3cm}
\label{tab:data}
\small\resizebox{4.930cm}{!}{%
\begin{tabular}{ccccc} 
\hline
Corpus      & Class        & Spk & Train\_utt & Test\_utt  \\ 
\hline
            & Minors     & 65   & 251        & 17         \\
VoxCeleb-    & Youngs     & 2693   & 12988      & 1466       \\
Enrichment & Middles    & 2583   & 18272      & 2086       \\
            & Seniors    & 1260   & 10389      & 1087       \\ 
\hline
            & Angry      & 4   & 1200       & 400        \\
            & Fear       & 4   & 1200       & 400        \\
CASIA       & Happy      & 4   & 1200       & 400        \\
            & Neutral    & 4   & 1200       & 400        \\
            & sad        & 4   & 1200       & 400        \\
            & surprise   & 4   & 1200       & 400        \\ 
\hline
            & Mandarin   & 24  & 7198       & 1800       \\
            & Cantonese  & 24  & 7559       & 1800       \\
            & Indonesian & 24  & 7671       & 1800       \\
            & Japanese   & 24  & 7662       & 1800       \\
AP17-OLR   & Russian    & 24  & 7190       & 1800       \\
            & Korean     & 24  & 7196       & 1800       \\
            & Vietnamese & 24  & 7200       & 1800       \\
            & Kazakh     & 86  & 4200       & 1800       \\
            & Tibetan    & 34  & 11100       & 1800       \\
            & Uyghur     & 353 & 5800       & 1800       \\
\hline
\end{tabular}}
\end{table}
\vspace{-0.2cm}
AP17-OLR consists of 10 different languages. The test set contains three subsets with different durations (1 second, 3 second, and full length). For speech emotion classification task, we conduct out experiments on CASIA with six emotion categories (i.e., angry, surprise, sad, fear, happy, and neutral). For the age group classification, we used the VoxCeleb Enrichment dataset to train the model. VoxCeleb Enrichment were extracted from YouTube videos, the audio clips were recorded in a variety of acoustic environments. The audios were divided into four age groups
.

\subsection{Experimental Settings}
 The input features are 32-dimensional Mel Filter-Banks extracted using the librosa package~\cite{mcfee2015librosa} with a window length of 25ms and a shift of 10ms with Hamming window. Mean and variance normalization is applied during instance normalization on Mel Filter-Banks features. A 192-frame segment (320-frame for age classification) is randomly chunked from each utterance.


All our experiments are based on the Wide-ResNet architecture~\cite{zagoruyko2016wide}, featuring a width of 5, depth of 28, payload learning rate of 0.2, 50 SGLD sampling steps, and a buffer size of 10,000
.
Frame-level feature extraction is based on ResNet topology with 3 groups of residual blocks.
Then the frame-level features are fed into the average pooling layer to get utterance-level embeddings, and the final classification layer dimensions are 10, 6, and 4, respectively, for language, emotion, and age group classification.
\begin{figure}[htbp]
\centering 
\includegraphics[width=0.8373\linewidth]{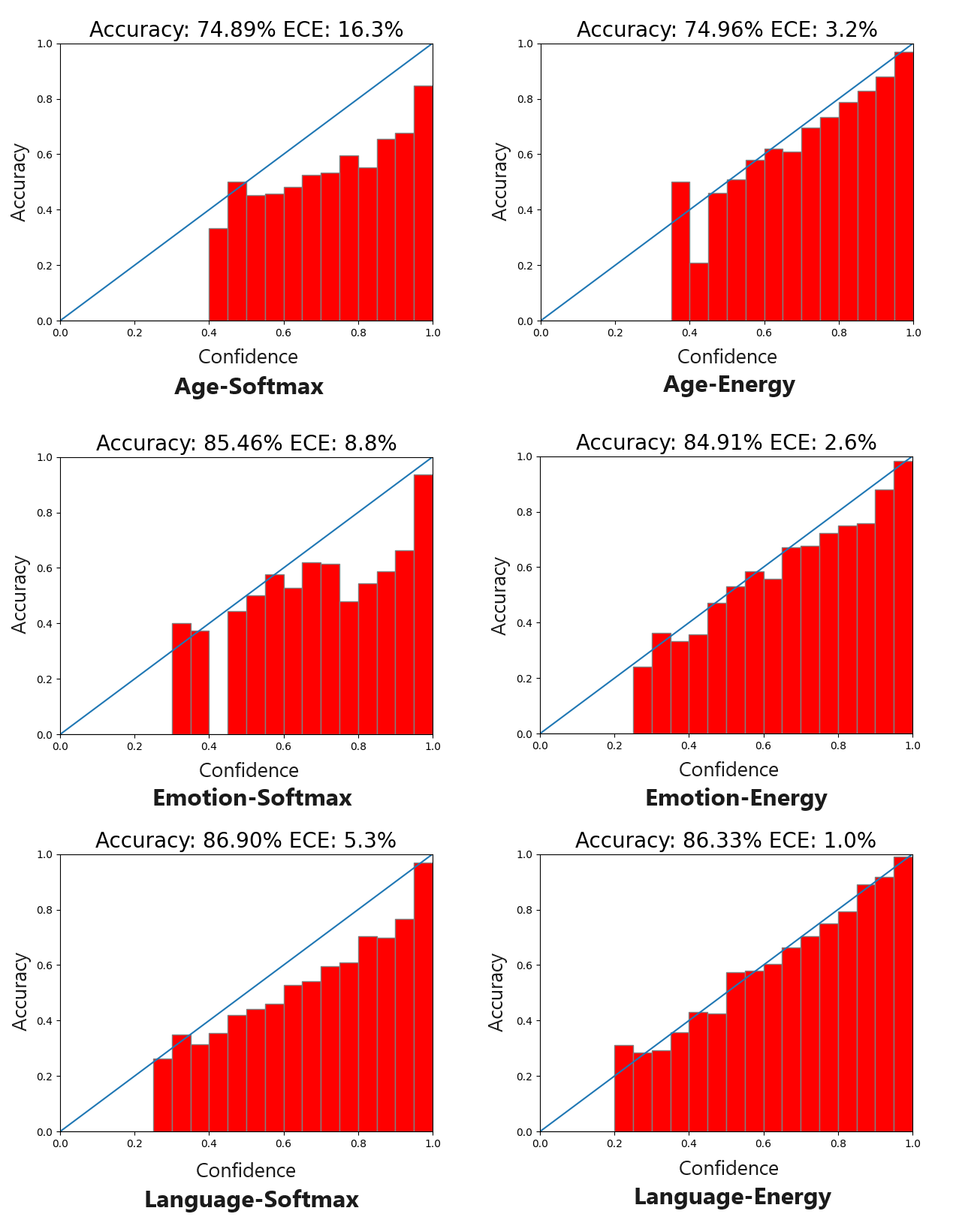} \vspace{-0.23cm}
\caption{Calibration results for three speech classification tasks. A smaller ECE indicates better calibration.} \vspace{-0.377cm}
\label{ECE} 
\end{figure}
We optimize the model with stochastic gradient descent (SGD)~\cite{bottou2010large} optimizer, the learning rate warms up to 0.1 during the first 1000 steps, and reduce the learning rate at epoch [40, 80, 120] with a decay rate of 0.2. Both softmax-based models and EBMs stick to the same training settings.
\subsection{Performance analysis and discussion} 
\begin{figure}[htbp]
\centering 
\includegraphics[width=0.893\linewidth,height=0.99\linewidth]{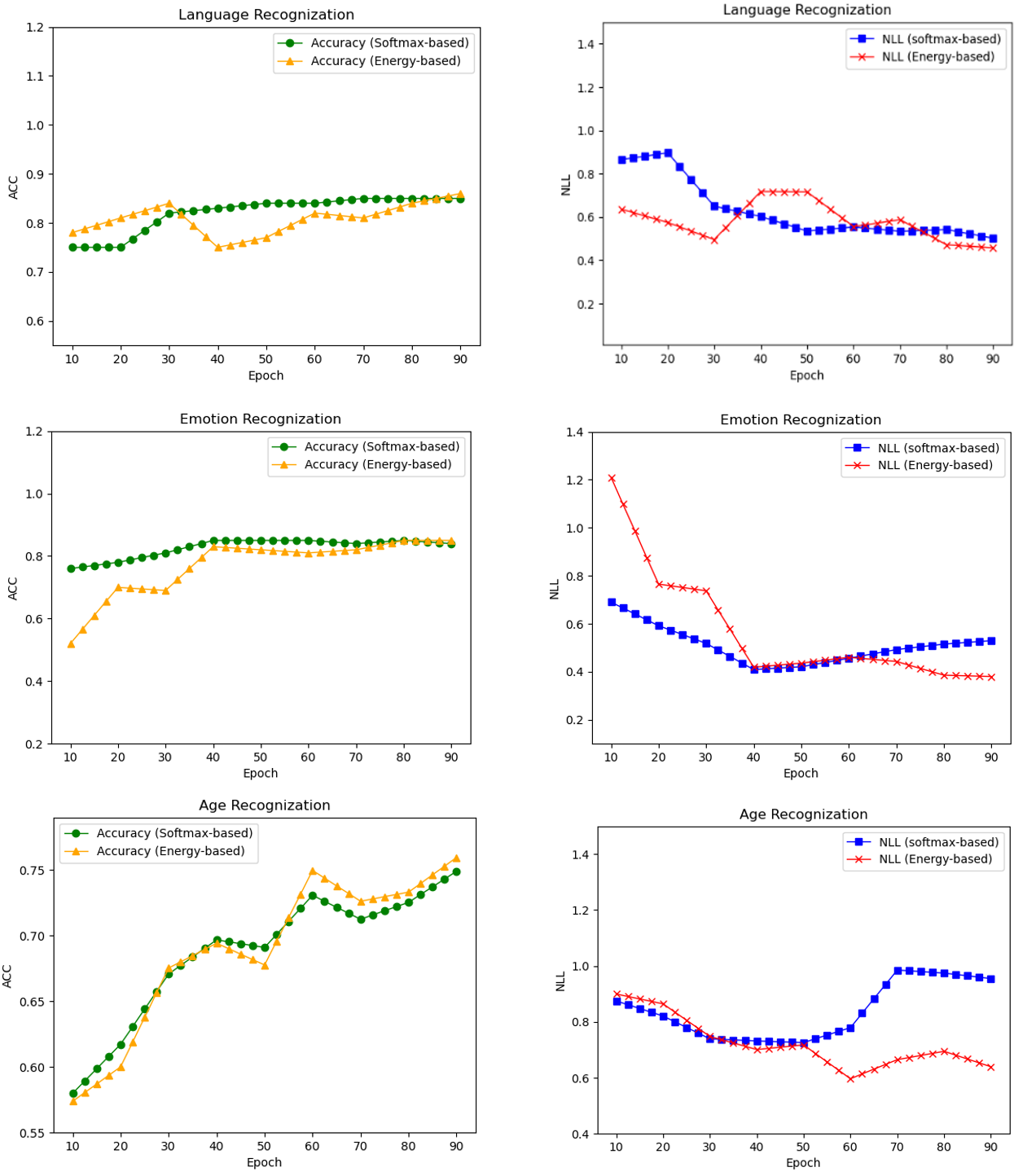} \vspace{-0.3cm}
\caption{The evolving trends of test ACC and NLL for three speech classification tasks with respect to training epochs.} \label{nll-acc}
\end{figure}

\vspace{-0.1cm}
\begin{table}[htbp]
    \centering
    \renewcommand{\arraystretch}{1.37} 
    \caption{Performance Comparison}
    \label{tab:performance}
    \resizebox{6.973cm}{!}{
    \begin{tabular}{lcccccc}
        \toprule
        \textbf{Task} &  \textbf{Method} & \multicolumn{3}{c}{ \textbf{Performance}} \\
        \cmidrule(lr){3-5}
        & &  \textbf{ECE(\%)} $\downarrow$ &  \textbf{ACC(\%)} $\uparrow$&  \textbf{NLL}$\downarrow$ \\
        \midrule
        Age & Softmax Classifier & 16.332 & 74.893  & 0.956\\
            (four classes)  & Energy-based Classifier &$\mathbf{3.208}$ & $\mathbf{74.960}$ &$\mathbf{0.596}$\\
        \midrule
        Emotion  & Softmax Classifier & 8.844 & $\mathbf{85.461}$ &0.516\\
                (six classes) & Energy-based Classifier & $\mathbf{2.620}$ & 84.910 &$\mathbf{0.381}$\\
        \midrule
         Language & Softmax Classifier & 5.314 & $\mathbf{86.900}$ &0.439\\
               (ten classes) & Energy-based Classifier & $\mathbf{1.044}$ & $86.334$ &$\mathbf{0.419}$\\
        \bottomrule
    \end{tabular}}
\end{table}\vspace{-0.273cm}
Table~\ref{tab:performance} summarizes the performance of softmax and energy-based classifiers across three classification tasks. The energy-based classifier outperform in the age classification task, achieving a higher accuracy of 74.96\% and significantly improving calibration with a reduction in ECE from 16.332\% to 3.208\%. While the emotion  classification task showed a minor accuracy drop, the EBMs exhibited substantial calibration improvement. Likewise, in the language  classification task, the energy model demonstrated superior calibration with a notable reduction in ECE. These results highlight the efficacy of energy-based classifiers in enhancing calibration across diverse tasks, suggesting their potential for reliable predictions with well-calibrated uncertainty estimates.
\\
\textbf{ Reliability diagrams.}
To evaluate calibration performance, reliability diagrams are utilized, visually representing the consistency between predicted probabilities and actual outcomes in Figures 1. It's evident that EBMs exhibit superior calibration, displaying smaller gaps and significantly lower  ECE compared to softmax-based models across three classification scenarios. Particularly, softmax-based models in three speech tasks consistently demonstrate overconfidence, as they tend to assign excessively high probabilities to predicted classes, a common issue observed in deep learning models \cite{guo2017calibration}. Notably, the reliability diagram for language classification using EBMs closely follows the diagonal, achieving an ECE of only 1.0\%, indicating nearly perfect calibration. These findings underscore that EBMs can significantly alleviate the issue of overconfidence in speech classification models, thereby achieving better calibration performance.
\vspace{-0.37cm}
\begin{table}[htbp]
\centering
\caption{Comparison of ECE (\%) for Different  Methods}\label{fig-temp}
\resizebox{6.37901cm}{!}{%
\begin{tabular}{cccc}
\hline
Methods & Age  &Emotion  &Language  \\
\hline
Softmax & 16.332 &8.844 & 5.314  \\
Temperature Scaling &15.027  & 7.965 & 2.297
 \\
Logistic Scaling &14.282   &6.809  &1.463    \\
Energy(Ours) & \textbf{3.208} &\textbf{2.620} &\textbf{1.044} \\
\hline
\end{tabular}}
\end{table}
\\
\textbf{Comparative evaluation of other calibration methods.}
We conduct a comparative analysis between EBMs and two post-processing calibration methods, namely Temperature Scaling and Logistic Scaling, across three speech classification tasks. The results are summarized in Table~\ref{fig-temp} alongside ECEs. It is evident that these two post-processing calibrators provide limited improvement in model calibration. This restricted effectiveness may stem from potential disparities between the auxiliary data and the target distribution, resulting in suboptimal calibration adjustments.
In contrast, EBMs consistently demonstrate superior performance, resulting in significant reductions in the ECE, without the requirement of supplementary training data.\\
\textbf{Why softmax-based models are poorly calibrated?}
As illustrated in Figure~\ref{nll-acc}, softmax models prioritize optimizing accuracy over achieving minimizing NLL. This observation aligns with the findings in \cite{guo2017calibration}, which suggest that modern neural networks can overfit to NLL without overfitting accuracy.
This phenomenon indicates that while softmax models may achieve high classification accuracy, they may not necessarily provide well-calibrated probability estimates. In other words, the pursuit of higher accuracy values can sometimes come at the cost of the model's ability to accurately reflect confidence in its predictions, thereby compromising calibration quality.
In contrast, EBMs excel in reducing NLL while maintaining high accuracy. Despite slower convergence during EBM training, they achieve lower NLL and higher confidence levels. 
\\
\textbf{Confidence distribution.} We analyze the problem of model overconfidence by visualizing the confidence distribution in Figure 3.  It demonstrates a significant prevalence of excessively high confidence levels for incorrect predictions across three speech tasks, resulting in unreliable confidence estimates for softmax-based models.  For example, within the confidence range of 0.9-1, the softmax-based model for age recognition yields 300 misclassified samples, whereas EBMs show only 10 misclassifications. It is noteworthy that EBMs exhibit a reduction in the confidence range of incorrect predictions across the three speech tasks, with accurate predictions predominantly falling within higher confidence intervals, leading to reliable confidence. 
\vspace{-0.2cm}
\vspace{-0.34cm}
\vspace{-0.317cm}
\begin{figure}[htbp]
\centering 
\includegraphics[width=0.8993\linewidth,height=0.997\linewidth]{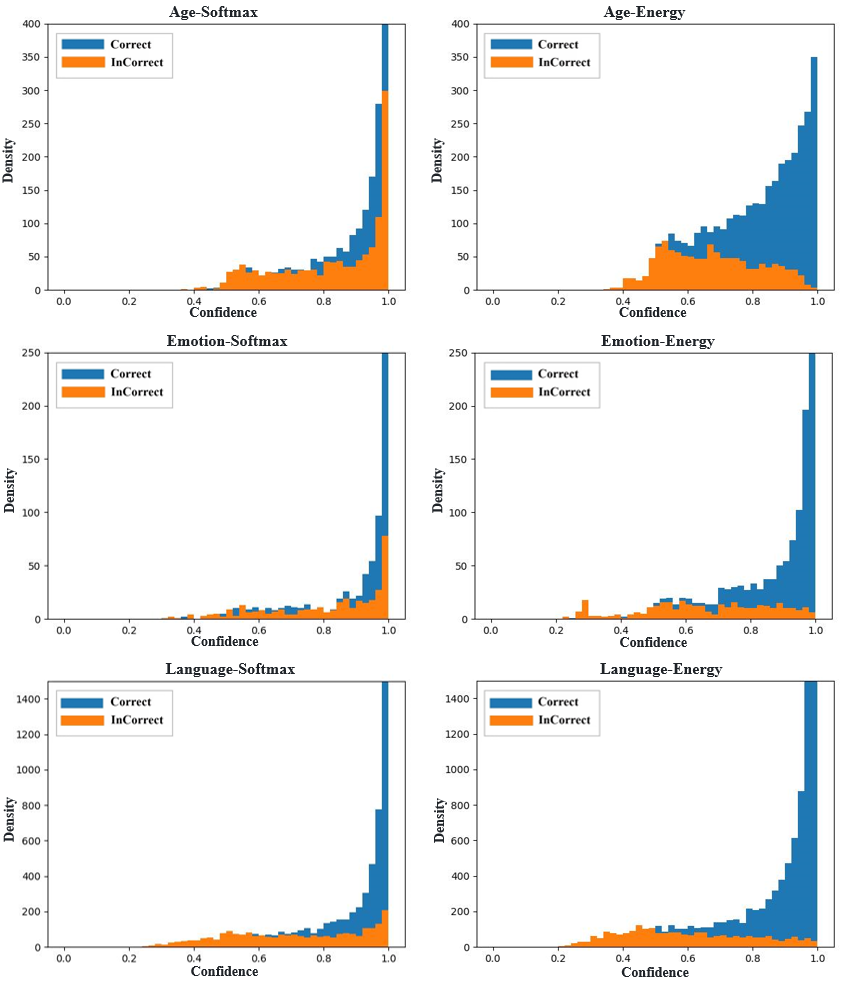} \vspace{-0.3cm}
\caption{Confidence distributions of softmax-based model and energy-based model.} \label{conf-dens}\vspace{-0.7cm}
\end{figure}
\section{Conclusions}
In this study, we explored the effectiveness of joint EBMs in calibrating speech classification tasks. Our results show that joint EBMs optimize both the classifier and the generative model to enhance calibration by gaining profound insights into the data distribution while also serving as a regularization mechanism, effectively mitigating overfitting tendencies.  These findings demonstrate that EBMs can notably generate well-calibrated predictions without compromising accuracy across diverse speech classification tasks.

\bibliographystyle{IEEEtran}
\bibliography{mybib}

\begin{thebibliography}{10}
\providecommand{\url}[1]{#1}
\csname url@samestyle\endcsname
\providecommand{\newblock}{\relax}
\providecommand{\bibinfo}[2]{#2}
\providecommand{\BIBentrySTDinterwordspacing}{\spaceskip=0pt\relax}
\providecommand{\BIBentryALTinterwordstretchfactor}{4}
\providecommand{\BIBentryALTinterwordspacing}{\spaceskip=\fontdimen2\font plus
\BIBentryALTinterwordstretchfactor\fontdimen3\font minus \fontdimen4\font\relax}
\providecommand{\BIBforeignlanguage}[2]{{%
\expandafter\ifx\csname l@#1\endcsname\relax
\typeout{** WARNING: IEEEtran.bst: No hyphenation pattern has been}%
\typeout{** loaded for the language `#1'. Using the pattern for}%
\typeout{** the default language instead.}%
\else
\language=\csname l@#1\endcsname
\fi
#2}}
\providecommand{\BIBdecl}{\relax}
\BIBdecl

\bibitem{rajendran2021language}
S.~Rajendran, S.~K. Mathivanan, P.~Jayagopal, M.~Venkatasen, T.~Pandi, M.~Sorakaya~Somanathan, M.~Thangaval, and P.~Mani, ``Language dialect based speech emotion recognition through deep learning techniques,'' \emph{International Journal of Speech Technology}, vol.~24, pp. 625--635, 2021.

\bibitem{yu2011calibration}
D.~Yu, J.~Li, and L.~Deng, ``Calibration of confidence measures in speech recognition,'' \emph{IEEE Transactions on Audio, Speech, and Language Processing}, vol.~19, no.~8, pp. 2461--2473, 2011.

\bibitem{guo2017calibration}
C.~Guo, G.~Pleiss, Y.~Sun, and K.~Q. Weinberger, ``On calibration of modern neural networks,'' in \emph{International conference on machine learning}.\hskip 1em plus 0.5em minus 0.4em\relax PMLR, 2017, pp. 1321--1330.

\bibitem{gitman2023confidence}
I.~Gitman, V.~Lavrukhin, A.~Laptev, and B.~Ginsburg, ``Confidence-based ensembles of end-to-end speech recognition models,'' in \emph{International conference on machine learning}.\hskip 1em plus 0.5em minus 0.4em\relax PMLR, 2017, pp. 11\,414--1418.

\bibitem{huang2021context}
S.~Huang, Y.~Luo, Z.~Zhuang, J.-G. Yu, M.~He, and Y.~Wang, ``Context-aware selective label smoothing for calibrating sequence recognition model,'' in \emph{Proceedings of the 29th ACM International Conference on Multimedia}, 2021, pp. 4591--4599.

\bibitem{chouimportance}
H.-C. Chou, L.~Goncalves, S.-G. Leem, C.-C. Lee, and C.~Busso, ``The importance of calibration: Rethinking confidence and performance of speech multi-label emotion classifiers,'' in \emph{Proc. {INTERSPEECH} 2023 -- 24\textsuperscript{rd} Annual Conference of the International Speech Communication Association}, {Dublin, Ireland}, {Aug.} 2023, p. 641–645.

\bibitem{kubik2022underconfidence}
V.~Kubik, A.~Jemstedt, H.~M. Eshratabadi, B.~L. Schwartz, and F.~U. J{\"o}nsson, ``The underconfidence-with-practice effect in action memory: The contribution of retrieval practice to metacognitive monitoring,'' \emph{Metacognition and Learning}, vol.~17, no.~2, pp. 375--398, 2022.

\bibitem{wang2021confident}
X.~Wang, H.~Liu, C.~Shi, and C.~Yang, ``Be confident! towards trustworthy graph neural networks via confidence calibration,'' \emph{Advances in Neural Information Processing Systems}, vol.~34, pp. 23\,768--23\,779, 2021.

\bibitem{wang2023calibrating}
D.~Wang, B.~Gong, and L.~Wang, ``On calibrating semantic segmentation models: Analyses and an algorithm,'' in \emph{Proceedings of the IEEE/CVF Conference on Computer Vision and Pattern Recognition}, 2023, pp. 23\,652--23\,662.

\bibitem{pereyra2017regularizing}
G.~Pereyra, G.~Tucker, J.~Chorowski, {\L}.~Kaiser, and G.~Hinton, ``Regularizing neural networks by penalizing confident output distributions,'' \emph{arXiv preprint arXiv:1701.06548}, 2017.

\bibitem{jung2020posterior}
T.~Jung, D.~Kang, H.~Cheng, L.~Mentch, and T.~Schaaf, ``Posterior calibrated training on sentence classification tasks,'' in \emph{Proceedings of the 58th Annual Meeting of the Association for Computational Linguistics}, 2020, pp. 2723--2730.

\bibitem{grathwohl2019your}
W.~Grathwohl, K.-C. Wang, J.-H. Jacobsen, D.~Duvenaud, M.~Norouzi, and K.~Swersky, ``Your classifier is secretly an energy based model and you should treat it like one,'' in \emph{International Conference on Learning Representations}, 2019.

\bibitem{song2021train}
Y.~Song and D.~P. Kingma, ``How to train your energy-based models,'' \emph{arXiv preprint arXiv:2101.03288}, 2021.

\bibitem{yang2021jem++}
X.~Yang and S.~Ji, ``Jem++: Improved techniques for training jem,'' in \emph{Proceedings of the IEEE/CVF International Conference on Computer Vision}, 2021, pp. 6494--6503.

\bibitem{yang2023towards}
X.~Yang, Q.~Su, and S.~Ji, ``Towards bridging the performance gaps of joint energy-based models,'' in \emph{Proceedings of the IEEE/CVF Conference on Computer Vision and Pattern Recognition}, 2023, pp. 15\,732--15\,741.

\bibitem{lecun2006tutorial}
Y.~LeCun, S.~Chopra, R.~Hadsell, M.~Ranzato, and F.~Huang, ``A tutorial on energy-based learning,'' \emph{Predicting structured data}, vol.~1, no.~0, 2006.

\bibitem{ou2024energy}
Z.~Ou \emph{et~al.}, ``Energy-based models with applications to speech and language processing,'' \emph{Foundations and Trends{\textregistered} in Signal Processing}, vol.~18, no. 1-2, pp. 1--199, 2024.

\bibitem{liu2020energy}
W.~Liu, X.~Wang, J.~Owens, and Y.~Li, ``Energy-based out-of-distribution detection,'' \emph{Advances in neural information processing systems}, vol.~33, pp. 21\,464--21\,475.

\bibitem{mortier2023calibration}
T.~Mortier, V.~Bengs, E.~H{\"u}llermeier, S.~Luca, and W.~Waegeman, ``On the calibration of probabilistic classifier sets,'' in \emph{International Conference on Artificial Intelligence and Statistics}.\hskip 1em plus 0.5em minus 0.4em\relax PMLR, 2023, pp. 8857--8870.

\bibitem{wang2016ap16}
D.~Wang, L.~Li, D.~Tang, and Q.~Chen, ``Ap16-ol7: A multilingual database for oriental languages and a language recognition baseline,'' in \emph{2016 Asia-Pacific Signal and Information Processing Association Annual Summit and Conference (APSIPA)}.\hskip 1em plus 0.5em minus 0.4em\relax IEEE, 2016, pp. 1--5.

\bibitem{li2017cheavd}
Y.~Li, J.~Tao, L.~Chao, W.~Bao, and Y.~Liu, ``Cheavd: a chinese natural emotional audio--visual database,'' \emph{Journal of Ambient Intelligence and Humanized Computing}, vol.~8, pp. 913--924, 2017.

\bibitem{hechmi2021voxceleb}
K.~Hechmi, T.~N. Trong, V.~Hautam{\"a}ki, and T.~Kinnunen, ``Voxceleb enrichment for age and gender recognition,'' in \emph{2021 IEEE Automatic Speech Recognition and Understanding Workshop (ASRU)}.\hskip 1em plus 0.5em minus 0.4em\relax IEEE, 2021, pp. 687--693.

\bibitem{mcfee2015librosa}
B.~McFee, C.~Raffel, D.~Liang, D.~P. Ellis, M.~McVicar, E.~Battenberg, and O.~Nieto, ``librosa: Audio and music signal analysis in python,'' in \emph{Proceedings of the 14th python in science conference}, vol.~8, 2015, pp. 18--25.

\bibitem{zagoruyko2016wide}
S.~Zagoruyko and N.~Komodakis, ``Wide residual networks,'' \emph{arXiv preprint arXiv:1605.07146}, 2016.

\bibitem{bottou2010large}
L.~Bottou, ``Large-scale machine learning with stochastic gradient descent,'' in \emph{Proceedings of COMPSTAT'2010: 19th International Conference on Computational StatisticsParis France, August 22-27, 2010 Keynote, Invited and Contributed Papers}.\hskip 1em plus 0.5em minus 0.4em\relax Springer, 2010, pp. 177--186.

\end{thebibliography}

\end{document}